\documentclass[preprint,12pt]{elsarticle}

\usepackage{amssymb}

\usepackage{dcolumn}

\usepackage{graphicx}

\biboptions{sort&compress}

\begin{document}

\begin{frontmatter}

\title{Group theoretical analysis of a quantum-mechanical three-dimensional quartic
anharmonic oscillator}

\author{Francisco M. Fern\'{a}ndez} \ead{fernande@quimica.unlp.edu.ar}

\address{INIFTA (UNLP, CCT La Plata-CONICET), Divisi\'{o}n Qu\'{i}mica Te\'{o}rica,
Diag. 113 y 64 (S/N), Sucursal 4, Casilla de Correo 16, 1900 La
Plata, Argentina}

\begin{abstract}
This paper illustrates the application of group theory to a
quantum-mechanical three-dimensional quartic anharmonic oscillator
with $O_{h}$ symmetry. It is shown that group theory predicts the
degeneracy of the energy levels and facilitates the application of
perturbation theory and the Rayleigh-Ritz variational method as
well as the interpretation of the results in terms of the symmetry
of the solutions . We show how to obtain suitable symmetry-adapted
basis sets.
\end{abstract}


\end{frontmatter}

\section{Introduction}

\label{sec:intro}

Quantum-mechanical anharmonic oscillators have proved useful for
the analysis of the vibration-rotation spectra of polyatomic
molecules\cite {C76,CSK78,CH86}. Several aspects of such spectra
as well as other molecular properties have been modelled by means
of simple coupled oscillators mainly with cubic and quartic
anharmonicities\cite
{P58,SS75,WH75,HSD79,NKM80,DH81,FF83,SDPWB84,SLN84,ZHBFR87,
S88,SC89,A94,NRRG96,RC96,CM97,SG98}. These and other applications
of the ubiquitous quantum-mechanical anharmonic oscillators
motivated their study and the development of suitable methods for
the treatment of the corresponding eigenvalue equations\cite
{BBW73a,BBW73b,EM74,SNR74,NR74,P74,H74,NM75,CGM76,DS77,NM77,B78,
HMM78,GR79,CGM79,DH79,NKTM80,
TB80,RBG80,WJ81,PE81a,PE81b,NKM81,MAFC82,CB82,QZLKL82,BEP82,L83,
UM84,DNSA84,NN84,RR84,AD85,CMW86,RZFH87,FH88,FE89,HZBR89,T89,AB91,
EC92,CSB93,W93,H94,SC95,DWW96,TE96a,TE96b,GS99,AF10}. The list of
papers just mentioned is far from being exhaustive and its purpose
is merely to give an idea of the interest aroused by the
anharmonic oscillators along the years.

Some authors have taken into account the symmetry of the anharmonic
oscillators in order to simplify their treatment\cite{T89,AB91,TE96b} and
others resorted to the more formal point group symmetry (PGS) to classify
the vibrational states\cite{T89,SC89,DH79,NKTM80,UM84,EC92}. Despite the
great amount of information provided by PGS none of those papers exhibits a
full application of such mathematical tool, except the analysis of
two-dimensional cubic and quartic oscillators by Pullen and Edmonds\cite
{PE81a,PE81b}. These papers motivated a recent application of PGS to a
variety of Hermitian\cite{AF09,F13b,FG14c,F14} and non-Hermitian anharmonic
oscillators with space-time symmetry\cite{FG14a, FG14b, AFG14a,AFG14b,AFG14c}%
. In the latter case PGS proved suitable for determining the conditions that
complex anharmonic potentials should satisfy in order to support real
eigenvalues.

The aim of this paper is to reinforce the idea that PGS is most
important in the study of quantum-mechanical anharmonic
oscillators because we deem that such technique has been
surprisingly overlooked or disregarded (except for the few works
already mentioned above\cite{PE81a,PE81b, AF09,F13b,FG14c,F14,
FG14a, FG14b, AFG14a,AFG14b,AFG14c}).

In Section~\ref{sec:model} we introduce the model, a quartic anharmonic
oscillator with considerably large symmetry described by the point group $%
O_{h}$. This problem was treated before by means of perturbation
theory and a less formal approach to symmetry based on parity and
coordinate-permutation operations\cite{T89}. Here we classify the
eigenstates of the unperturbed Hamiltonian according to the
irreducible representations (irreps) of that point group and
predict the rupture of the degeneracy by the quartic perturbation.
In Section~\ref{sec:perturbation} we apply perturbation theory
through second order to verify the splitting of the eigenspaces
predicted in the preceding section. In Section~\ref
{sec:Rayleigh-Ritz} we discuss the application of the
Rayleigh-Ritz variational method with basis sets adapted to the
symmetry of the problem. In three consecutive subsections we
discuss the harmonic oscillator basis set, the Krylov space and a
non-orthogonal basis set. We show results illustrating the
splitting of the unperturbed eigenspaces due to the quartic
perturbation that breaks the symmetry of the system. In
Section~\ref{sec:conclusions} we summarize the main results of the
paper and draw conclusions. Finally, in the \ref{sec:appendix} we
outline the main features of group theory that are necessary for
the analysis of the present problem.

\section{Model}

\label{sec:model}

Among the many models mentioned above we have chosen a three-dimensional
quartic anharmonic oscillator already studied earlier by Turbiner\cite{T89}
\begin{eqnarray}
H &=&p_{x}^{2}+p_{y}^{2}+p_{z}^{2}+x^{2}+y^{2}+z^{2}+\lambda \left[ \beta
\left( x^{4}+y^{4}+z^{4}\right) +x^{2}y^{2}+x^{2}z^{2}+y^{2}z^{2}\right] ,\;
\nonumber \\
&&\lambda >0,\;\beta >0,  \label{eq:H}
\end{eqnarray}
who recognized that it exhibits the symmetry of a cube. The author
stated that his classification based on parity and permutation
operators was incomplete. In this paper we describe the symmetry
properties of this model by means of the point group
$O_{h}$\cite{H62,T64,C90}. In other words, this Hamiltonian
operator is invariant with respect to the symmetry operations
indicated in the table of characters shown in Table~\ref{tab:Oh}
described in the \ref{sec:appendix}. A detailed discussion of the
construction of the matrix representation of the symmetry
operations for the $O_{h}$ point group is available
elsewhere\cite{F14}.

The eigenvalues $E_{k\,m\,n}^{(0)}$ and eigenfunctions $\varphi
_{k\,m\,n}(x,y,z)$ of $H_{0}=H(\lambda =0)$ are
\begin{eqnarray}
E_{k\,m\,n}^{(0)} &=&2(k+m+n)+3  \nonumber \\
\varphi _{k\,m\,n}(x,y,z) &=&\phi _{k}(x)\phi _{m}(y)\phi
_{n}(z),\;k,m,n=0,1,\ldots ,  \label{eq:eigenv_eigenf_H0}
\end{eqnarray}
where $\phi _{j}(q)$ is an eigenfunction of the one-dimensional harmonic
oscillator $H_{HO}=p_{q}^{2}+q^{2}$. Every energy level is $\frac{(\nu
+1)(\nu +2)}{2}$-fold degenerate, where $\nu =k+m+n$.

Throughout this paper we resort to the following notation for the
permutation of a set of three real numbers
\begin{eqnarray}
\{a,a,a\}_{P} &=&\{a,a,a\}  \nonumber \\
\{a,b,b\}_{P} &=&\{\{a,b,b\},\{b,a,b\},\{b,b,a\}\}  \nonumber \\
\{a,b,c\}_{P}
&=&\{\{a,b,c\},\{c,a,b\},\{b,c,a\},\{b,a,c\},\{c,b,a\},\{a,c,b\}\},
\label{eq:permutations}
\end{eqnarray}
where it is assumed that $a\neq b$, $a\neq c$, and $b\neq c$. It enables us
to express the symmetry of the unperturbed eigenfunctions ($\lambda =0$) as
\begin{equation}
\begin{array}{ll}
\{2n,2n,2n\} & A_{1g} \\
\{2n+1,2n+1,2n+1\} & A_{2u} \\
\{2n+1,2n+1,2m\}_{P} & T_{2g} \\
\{2n,2n,2m+1\}_{P} & T_{1u} \\
\{2n,2n,2m\}_{P} & A_{1g},E_{g} \\
\{2n+1,2n+1,2m+1\}_{P} & A_{2u},E_{u} \\
\{2n,2m,2k\}_{P} & A_{1g},A_{2g},E_{g},E_{g} \\
\{2n+1,2m+1,2k+1\}_{P} & A_{1u},A_{2u},E_{u},E_{u} \\
\{2n,2m,2k+1\}_{P} & T_{1u},T_{2u} \\
\{2n+1,2m+1,2k\}_{P} & T_{1g},T_{2g}
\end{array}
.  \label{eq:psi_0_symmetry}
\end{equation}

The anharmonic part of the potential reduces the symmetry of the
system and the degeneracy of the energy levels is consequently
diminished when $\lambda >0$ causing a splitting of the energy
levels. For the lowest ones this splitting is given by
\begin{eqnarray}
\nu &=&0\rightarrow A_{1g}  \nonumber \\
\nu &=&1\rightarrow T_{1u}  \nonumber \\
\nu &=&2\rightarrow A_{1g},E_{g},T_{2g}  \nonumber \\
\nu &=&3\rightarrow A_{2u},T_{1u},T_{1u},T_{2u}  \nonumber \\
\nu &=&4\rightarrow A_{1g},A_{1g},E_{g},E_{g},T_{1g},T_{2g},T_{2g}.
\label{eq:splitting}
\end{eqnarray}
This equation tells us that the lowest energy level is nondegenerate, the
first excited energy level remains three-fold degenerate, the second excited
energy level splits into a singlet a doublet and a triplet, the third
excited energy level splits into a singlet and three triples (two of them of
the same symmetry $T_{1u}$), etc. Note the alternating parity (either $g$ or
$u$) given by $(-1)^{\nu }$.

For simplicity, in this paper we restrict ourselves to the case $\beta =0$.
This choice reduces the number of parameters in the potential-energy
function but alters neither the symmetry of the problem nor the main
conclusions drawn from it.

\section{Perturbation theory}

\label{sec:perturbation}

The purpose of this section is merely to carry out a simple calculation
based on perturbation theory in order to verify the splitting of the energy
levels outlined by equation (\ref{eq:splitting}). There are several
strategies for obtaining the perturbation corrections when $H_{0}$ exhibits
degenerate states. Here we simply obtain the first two perturbation
corrections by inserting the truncated expansion $E=E^{(0)}+E^{(1)}\lambda
+E^{(2)}\lambda ^{2}$ into the characteristic polynomial given by the
secular determinant $|H-EI|=0$ and then solving the resulting equation for $%
E^{(1)}$ and $E^{(2)}$. This apparently impractical brute-force
approach is sufficient for present purposes; the only subtlety
being that each irrep is treated separately as discussed in
section~\ref{sec:Rayleigh-Ritz} (in particular with the basis set
of subsection~\ref{subsec:non-orth_basis}). For the first energy
levels we obtain the following results
\begin{equation}
E_{1A_{1g}}=3+\frac{3}{4}\lambda -\frac{15}{32}\lambda ^{2}+O(\lambda ^{3}),
\label{eq:PT_nu=0}
\end{equation}
\begin{equation}
E_{1T_{1u}}=5+\frac{7}{4}\lambda -\frac{51}{32}\lambda ^{2}+O(\lambda ^{3}),
\label{eq:PT_nu=1}
\end{equation}
\begin{eqnarray}
E_{1E_{g}} &=&7+\frac{9}{4}\lambda -\frac{9}{4}\lambda ^{2}+O(\lambda ^{3}),
\nonumber \\
E_{2A_{1g}} &=&7+\frac{15}{4}\lambda -\frac{171}{32}\lambda ^{2}+O(\lambda
^{3}),  \nonumber \\
E_{1T_{2g}} &=&7+\frac{15}{4}\lambda -\frac{147}{32}\lambda ^{2}+O(\lambda
^{3}),  \label{eq:PT_nu=2}
\end{eqnarray}
\begin{eqnarray}
E_{2T_{1u}} &=&9+\frac{13}{4}\lambda -\frac{225}{56}\lambda ^{2}+O(\lambda
^{3}),  \nonumber \\
E_{1T_{2u}} &=&9+\frac{21}{4}\lambda -\frac{57}{8}\lambda ^{2}+O(\lambda
^{3}),  \nonumber \\
E_{3T_{1u}} &=&9+\frac{27}{4}\lambda -\frac{2649}{224}\lambda ^{2}+O(\lambda
^{3}),  \nonumber \\
E_{1A_{2u}} &=&9+\frac{27}{4}\lambda -\frac{351}{32}\lambda ^{2}+O(\lambda
^{3}),  \label{eq:PT_nu=3}
\end{eqnarray}
which are consistent with the prediction of group theory shown in equation (%
\ref{eq:splitting}). It is worth noting that the degeneracy of the
pairs of energy levels $(E_{2A_{1g}},E_{1T_{2g}})$ and
$(E_{1A_{2u}},E_{3T_{1u}})$ breaks at second order (the correction
of first order being identical).\ However, such states offer no
difficulty because we can treat each member of the pair separate
from the other because they belong to different symmetry species.
This strategy is one of the advantages of a possible systematic
application of group theory to perturbation theory.

\section{Rayleigh-Ritz variational method}

\label{sec:Rayleigh-Ritz}

This approach is based on a variational ansatz given by a finite linear
combination of functions of a basis set $B=\left\{ f_{0},f_{1},\ldots
\right\} $:
\begin{equation}
\psi =\sum_{i=0}^{N-1}c_{i}f_{i}.  \label{eq:psi_ansatz}
\end{equation}
The variational coefficients $c_{i}$ are chosen so that the variational
integral
\begin{equation}
E=\frac{\left\langle \psi \right| H\left| \psi \right\rangle }{\left\langle
\psi \right. \left| \psi \right\rangle },  \label{eq:var_int}
\end{equation}
is a minimum. The condition $\partial E/\partial c_{j}=0$ leads to the
secular equations
\begin{equation}
\left\langle f_{j}\right| H-E\left| \psi \right\rangle =0,\;j=0,1,\ldots
,N-1,  \label{eq:secular_gen}
\end{equation}
from which we obtain the coefficients $c_{i}$ and approximate energies $%
E_{j}(N)$, $j=0,1,\ldots ,N-1$. This result is valid for
orthogonal as well as non-orthogonal basis functions $f_{j}$,
provided that in the latter case the functions are linearly
independent. This approach always yields increasingly tighter
upper bounds because $E_{j}(N)>E_{j}(N+1)$\cite{M33}.

If the system exhibits symmetry it is convenient to choose a
suitable basis set $B^{S}=\left\{ f_{0}^{S},f_{1}^{S},\ldots
\right\} $ for each irrep $S$. In order to obtain it we apply the
projection operator $P^{S}$ (see the \ref{sec:appendix} for more
details) to every $f_{i}\in B$ and then remove the linearly
dependent functions from the resulting set $\left\{
P^{S}f_{i},\;i=0,1,\ldots \right\} $. One advantage of using the basis sets $%
B^{S}$ is that the dimension of the secular equations for a given accuracy
is considerably smaller. Thus, instead of equations (\ref{eq:psi_ansatz})
and (\ref{eq:secular_gen}) we have
\begin{equation}
\psi ^{S}=\sum_{i=0}^{N_{S}-1}c_{i}^{S}f_{i}^{S},  \label{eq:psi_ansatz_S}
\end{equation}
and
\begin{equation}
\left\langle f_{j}^{S}\right| H-E^{S}\left| \psi ^{S}\right\rangle
=0,\;j=0,1,\ldots ,N_{S}-1,  \label{eq:sec_gen_S}
\end{equation}
for each irrep $S$. In this case we obtain upper bounds for each irrep
exactly in the same way as before: $E_{j}^{S}(N_{S})>E_{j}^{S}(N_{S}+1)$.

\subsection{Basis set of eigenfunctions of $H_{0}$}

\label{subsec:basis_H0}

One of the most convenient basis sets consists of linear combinations of
eigenfunctions of $H_{0}$ adapted to the symmetry of $H$. In this case the
variational method reduces to diagonalizing the matrix representation of the
Hamiltonian $\mathbf{H}^{S}$ in each orthonormal basis set $B^{S}=\{\varphi
_{k\,m\,n}^{S}\}$ adapted to the corresponding symmetry species $S$. We thus
obtain approximate eigenvalues $E^{S}$ and eigenfunctions $\psi ^{S}$ that
are linear combinations of the form
\begin{equation}
\psi ^{S}=\sum_{k,m,n}c_{k\,m\,n}^{S}\varphi _{k\,m\,n}^{S}.
\label{eq:psi^S}
\end{equation}
The coefficients $c_{k\,m\,n}^{S}$ and the approximate eigenvalues $E^{S}$
are given by the eigenvectors and eigenvalues of $\mathbf{H}^{S}$,
respectively, and are also solutions of a secular equation similar to (\ref
{eq:sec_gen_S}). The projection operators enable us to construct the
symmetry-adapted basis sets $B^{S}$ in the following way
\begin{equation}
P^{S}\varphi _{k\,m\,n}=\sum_{k^{\prime },m^{\prime },n^{\prime
}}u_{k\,m\,n}^{k^{\prime }\,m^{\prime }n^{\prime }}\varphi _{k^{\prime
}\,m^{\prime }\,n^{\prime }}.  \label{eq:P^S_phi_kmn}
\end{equation}
Since $[H_{0},P^{S}]=0$ then $u_{k\,m\,n}^{k^{\prime }\,m^{\prime }n^{\prime
}}=0$ unless $k+m+n=k^{\prime }+m^{\prime }+n^{\prime }=\nu $.

After removing the linearly dependent functions and orthonormalizing the
remaining ones the basis sets adapted to the symmetry of the Hamiltonian
operator (\ref{eq:H}) result to be
\begin{eqnarray}
A_{1g} &:&  \nonumber \\
&&\varphi _{2n\,2n\,2n}  \nonumber \\
&&\frac{1}{\sqrt{3}}\left( \varphi _{2n\,2m\,2m}+\varphi
_{2m\,2n\,2m}+\varphi _{2m\,2m\,2n}\right)  \nonumber \\
&&\frac{1}{\sqrt{6}}\left( \varphi _{2k\,2m\,2n}+\varphi
_{2n\,2k\,2m}+\varphi _{2m\,2n\,2k}+\varphi _{2m\,2k\,2n}+\varphi
_{2n\,2m\,2k}\right.  \nonumber \\
&&\left. +\varphi _{2k\,2n\,2m}\right) ,  \label{eq:phi_A1g}
\end{eqnarray}
\begin{eqnarray}
A_{2g} &:&  \nonumber \\
&&\frac{1}{\sqrt{6}}\left( \varphi _{2k\,2m\,2n}+\varphi
_{2n\,2k\,2m}+\varphi _{2m\,2n\,2k}-\varphi _{2m\,2k\,2n}-\varphi
_{2n\,2m\,2k}\right.  \nonumber \\
&&\left. -\varphi _{2k\,2n\,2m}\right) ,  \label{eq:phi_A2g}
\end{eqnarray}
\begin{eqnarray}
E_{g} &:&  \nonumber \\
&&\left\{ \frac{1}{\sqrt{6}}\left( 2\varphi _{2n\,2m\,2m}-\varphi
_{2m\,2n\,2m}-\varphi _{2m\,2m\,2n}\right) ,\frac{1}{\sqrt{2}}\left( \varphi
_{2m\,2n\,2m}-\varphi _{2m\,2m\,2n}\right) \right\}  \nonumber \\
&&\left\{ \frac{1}{\sqrt{6}}\left( 2\varphi _{2k\,2m\,2n}-\varphi
_{2n\,2k\,2m}-\varphi _{2m\,2n\,2k}\right) ,\frac{1}{\sqrt{2}}\left( \varphi
_{2n\,2k\,2m}-\varphi _{2m\,2n\,2k}\right) \right\}  \nonumber \\
&&\left\{ \frac{1}{\sqrt{6}}\left( 2\varphi _{2m\,2k\,2n}-\varphi
_{2n\,2m\,2k}-\varphi _{2k\,2n\,2m}\right) ,\frac{1}{\sqrt{2}}\left( \varphi
_{2n\,2m\,2k}-\varphi _{2k\,2n\,2m}\right) \right\} ,  \label{eq:phi_Eg}
\end{eqnarray}
\begin{eqnarray}
T_{1g} &:&  \nonumber \\
&&\left\{ \frac{1}{\sqrt{2}}\left( \varphi _{2k\,2m+1\,2n+1}-\varphi
_{2k\,2n+1\,2m+1}\right) ,\frac{1}{\sqrt{2}}\left( \varphi
_{2m+1\,2k\,2n+1}-\varphi _{2n+1\,2k\,2m+1}\right) ,\right.  \nonumber \\
&&\left. \frac{1}{\sqrt{2}}\left( \varphi _{2m+1\,2n+1\,2k}-\varphi
_{2n+1\,2m+1\,2k}\right) \right\} ,  \label{eq:phi_T1g}
\end{eqnarray}
\begin{eqnarray}
T_{2g} &:&  \nonumber \\
&&\left\{ \varphi _{2k\,2m+1\,2m+1},\varphi _{2m+1\,2k\,2m+1},\varphi
_{2m+1\,2m+1\,2k}\right\}  \nonumber \\
&&\left\{ \frac{1}{\sqrt{2}}\left( \varphi _{2k\,2m+1\,2n+1}+\varphi
_{2k\,2n+1\,2m+1}\right) ,\frac{1}{\sqrt{2}}\left( \varphi
_{2m+1\,2k\,2n+1}+\varphi _{2n+1\,2k\,2m+1}\right) ,\right.  \nonumber \\
&&\left. \frac{1}{\sqrt{2}}\left( \varphi _{2m+1\,2n+1\,2k}+\varphi
_{2n+1\,2m+1\,2k}\right) \right\} ,  \label{eq:phi_T2g}
\end{eqnarray}
\begin{eqnarray}
A_{1u} &:&  \nonumber \\
&&\frac{1}{\sqrt{6}}\left( \varphi _{2k+1\,2m+1\,2n+1}+\varphi
_{2n+1\,2k+1\,2m+1}+\varphi _{2m+1\,2n+1\,2k+1}-\varphi
_{2m+1\,2k+1\,2n+1}\right.  \nonumber \\
&&\left. -\varphi _{2n+1\,2m+1\,2k+1}-\varphi _{2k+1\,2n+1\,2m+1}\right) ,
\label{eq:phi_A1u}
\end{eqnarray}
\begin{eqnarray}
A_{2u} &:&  \nonumber \\
&&\varphi _{2n+1\,2n+1\,2n+1}  \nonumber \\
&&\frac{1}{\sqrt{6}}\left( \varphi _{2k+1\,2m+1\,2n+1}+\varphi
_{2n+1\,2k+1\,2m+1}+\varphi _{2m+1\,2n+1\,2k+1}+\varphi
_{2m+1\,2k+1\,2n+1}\right.  \nonumber \\
&&\left. +\varphi _{2n+1\,2m+1\,2k+1}+\varphi _{2k+1\,2n+1\,2m+1}\right) ,
\label{eq:phi_A2u}
\end{eqnarray}
\begin{eqnarray}
E_{u} &:&  \nonumber \\
&&\left\{ \frac{1}{\sqrt{6}}\left( 2\varphi _{2n+1\,2m+1\,2m+1}-\varphi
_{2m+1\,2n+1\,2m+1}-\varphi _{2m+1\,2m+1\,2n+1}\right) ,\right.  \nonumber \\
&&\left. \frac{1}{\sqrt{2}}\left( \varphi _{2m+1\,2n+1\,2m+1}-\varphi
_{2m+1\,2m+1\,2n+1}\right) \right\}  \nonumber \\
&&\left\{ \frac{1}{\sqrt{6}}\left( 2\varphi _{2k+1\,2m+1\,2n+1}-\varphi
_{2n+1\,2k+1\,2m+1}-\varphi _{2m+1\,2n+1\,2k+1}\right) ,\right.  \nonumber \\
&&\left. \frac{1}{\sqrt{2}}\left( \varphi _{2n+1\,2k+1\,2m+1}-\varphi
_{2m+1\,2n+1\,2k+1}\right) \right\} ,  \nonumber \\
&&\left\{ \frac{1}{\sqrt{6}}\left( 2\varphi _{2m+1\,2k+1\,2n+1}-\varphi
_{2n+1\,2m+1\,2k+1}-\varphi _{2k+1\,2n+1\,2m+1}\right) ,\right.  \nonumber \\
&&\left. \frac{1}{\sqrt{2}}\left( \varphi _{2n+1\,2m+1\,2k+1}-\varphi
_{2k+1\,2n+1\,2m+1}\right) \right\} ,  \label{eq:phi_Eu}
\end{eqnarray}
\begin{eqnarray}
T_{1u} &:&  \nonumber \\
&&\left\{ \varphi _{2k+1\,2m\,2m},\varphi _{2m\,2k+1\,2m},\varphi
_{2m\,2m\,2k+1}\right\}  \nonumber \\
&&\left\{ \frac{1}{\sqrt{2}}\left( \varphi _{2k+1\,2m\,2n}+\varphi
_{2k+1\,2n\,2m}\right) ,\frac{1}{\sqrt{2}}\left( \varphi
_{2m\,2k+1\,2n}+\varphi _{2n\,2k+1\,2m}\right) ,\right.  \nonumber \\
&&\left. \frac{1}{\sqrt{2}}\left( \varphi _{2m\,2n\,2k+1}+\varphi
_{2n\,2m\,2k+1}\right) \right\}  \label{eq:phi_T1u}
\end{eqnarray}
\begin{eqnarray}
T_{2u} &:&  \nonumber \\
&&\left\{ \frac{1}{\sqrt{2}}\left( \varphi _{2k+1\,2m\,2n}-\varphi
_{2k+1\,2n\,2m}\right) ,\frac{1}{\sqrt{2}}\left( \varphi
_{2m\,2k+1\,2n}-\varphi _{2n\,2k+1\,2m}\right) ,\right.  \nonumber \\
&&\left. \frac{1}{\sqrt{2}}\left( \varphi _{2m\,2n\,2k+1}-\varphi
_{2n\,2m\,2k+1}\right) \right\} .  \label{eq:phi_T2u}
\end{eqnarray}
By convention $\varphi _{i\,j\,k}$ means that all the subscripts are
different, equal subscripts are indicated explicitly as, for example, in $%
\varphi _{i\,j\,j}$.

\subsection{Krylov space}

\label{subsec:Krylov}

A particular basis set that spans what is commonly called the
Krylov space is given by $f_{i}=H^{i}f$, where $f$ is a suitably
chosen function. If follows from the properties of the projection
operators discussed in the \ref{sec:appendix} that
$P^{S}f_{i}=H^{i}P^{S}f$ $=H^{i}f^{S}$ so that by simply choosing
a seed function $f^{S}$ with the correct symmetry then the
resulting
basis set is automatically adapted to the corresponding irrep. We thus have $%
B^{S}=\left\{ f_{i}^{S}=H^{i}f^{S},\;i=0,1,\ldots \right\} $ for each irrep $%
S$ and solve secular equations similar to (\ref{eq:sec_gen_S}).

Suitable seed functions are
\begin{equation}
f^{A_{1g}}=\exp \left[ -a\left( x^{2}+y^{2}+z^{2}\right) \right] ,
\label{eq:f_A1g}
\end{equation}
\begin{equation}
f^{A_{2g}}=\left(
x^{4}y^{2}-x^{4}z^{2}-x^{2}y^{4}+x^{2}z^{4}+y^{4}z^{2}-y^{2}z^{4}\right)
\exp \left[ -a\left( x^{2}+y^{2}+z^{2}\right) \right] ,  \label{eq:f_A2g}
\end{equation}
\begin{equation}
f^{E_{g}}=\left\{
\begin{array}{c}
\left( 2z^{2}-x^{2}-y^{2}\right) \exp \left[ -a\left(
x^{2}+y^{2}+z^{2}\right) \right] \\
\left( x^{2}-y^{2}\right) \exp \left[ -a\left( x^{2}+y^{2}+z^{2}\right)
\right]
\end{array}
\right. ,  \label{eq:f_Eg}
\end{equation}
\begin{equation}
f^{T_{1g}}=\left\{
\begin{array}{c}
\left( xy^{3}-x^{3}y\right) \exp \left[ -a\left( x^{2}+y^{2}+z^{2}\right)
\right] \\
\left( xz^{3}-x^{3}z\right) \exp \left[ -a\left( x^{2}+y^{2}+z^{2}\right)
\right] \\
\left( yz^{3}-y^{3}z\right) \exp \left[ -a\left( x^{2}+y^{2}+z^{2}\right)
\right]
\end{array}
\right. ,  \label{eq:f_T1g}
\end{equation}
\begin{equation}
f^{T_{2g}}=\left\{
\begin{array}{c}
xy\exp \left[ -a\left( x^{2}+y^{2}+z^{2}\right) \right] \\
xz\exp \left[ -a\left( x^{2}+y^{2}+z^{2}\right) \right] \\
yz\exp \left[ -a\left( x^{2}+y^{2}+z^{2}\right) \right]
\end{array}
\right. ,  \label{eq:f_T2g}
\end{equation}
\begin{equation}
f^{A_{1u}}=\left(
x^{5}yz^{3}-x^{5}y^{3}z+x^{3}y^{5}z-x^{3}yz^{5}-xy^{5}z^{3}+xy^{3}z^{5}%
\right) \exp \left[ -a\left( x^{2}+y^{2}+z^{2}\right) \right] ,
\label{eq:f_A1u}
\end{equation}
\begin{equation}
f^{A_{2u}}=xyz\exp \left[ -a\left( x^{2}+y^{2}+z^{2}\right) \right] ,
\label{eq:f_A2u}
\end{equation}
\begin{equation}
f^{E_{u}}=\left\{
\begin{array}{c}
xyz\left( 2z^{2}-x^{2}-y^{2}\right) \exp \left[ -a\left(
x^{2}+y^{2}+z^{2}\right) \right] \\
xyz\left( x^{2}-y^{2}\right) \exp \left[ -a\left( x^{2}+y^{2}+z^{2}\right)
\right]
\end{array}
\right. ,  \label{eq:f_Eu}
\end{equation}
\begin{equation}
f^{T_{1u}}=\left\{
\begin{array}{c}
x\exp \left[ -a\left( x^{2}+y^{2}+z^{2}\right) \right] \\
y\exp \left[ -a\left( x^{2}+y^{2}+z^{2}\right) \right] \\
z\exp \left[ -a\left( x^{2}+y^{2}+z^{2}\right) \right]
\end{array}
\right. ,  \label{eq:f_T1u}
\end{equation}
\begin{equation}
f^{T_{2u}}=\left\{
\begin{array}{c}
x\left( y^{2}-z^{2}\right) \exp \left[ -a\left( x^{2}+y^{2}+z^{2}\right)
\right] \\
y\left( z^{2}-x^{2}\right) \exp \left[ -a\left( x^{2}+y^{2}+z^{2}\right)
\right] \\
z\left( x^{2}-y^{2}\right) \exp \left[ -a\left( x^{2}+y^{2}+z^{2}\right)
\right]
\end{array}
\right. ,  \label{eq:f_T2u}
\end{equation}
where $a>0$ is a nonlinear variational parameter that enables us to improve
the accuracy of the results.

We carried out a set of calculations with $a=1$ and encountered a
most surprising difficulty. For some reason (unknown to us at
present) the Rayleigh-Ritz variational method in the Krylov space
yields only one of the two $T_{1u}$ state triplets stemming from
$E(\lambda =0)=7$ and only one of the $T_{2g}$ state triplets
stemming from $E(\lambda =0)=11$. We do not investigate this fact
any further in this paper and just mention it in passing. This
fact is surprising because we applied this approach in the past to
other anharmonic oscillators and faced no such problem\cite{AF09,
FG14c}.

\subsection{Non-orthogonal basis set}

$\label{subsec:non-orth_basis}$

\begin{figure}[tbp]
\begin{center}
\includegraphics[width=12cm]{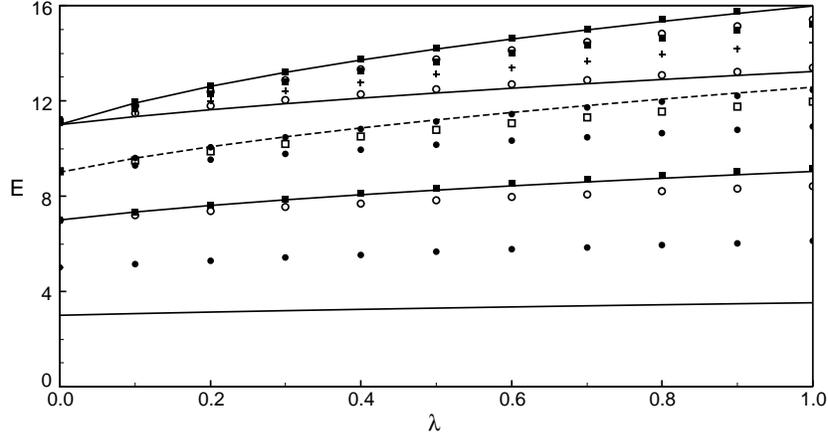} \
\par
\end{center}
\caption{First eigenvalues of the Hamiltonian operator $%
H=p_x^2+p_y^2+p_z^2+x^2+y^2+z^2+\lambda \left(x^2 y^2+x^2 z^2+y^2
z^2
\right) $ with symmetry $A_{1g}$ (solid line), $T_{1u}$ (filled circles), $%
E_g $ (empty circles), $T_{2g}$ (filled squares), $A_{2u}$ (dashed line), $%
T_{2u}$ (empty squares), $T_{1g}$ (crosses). } \label{fig:RROH}
\end{figure}

In addition to the orthonormal basis set
(\ref{eq:phi_A1g}-\ref{eq:phi_T2u}) we can also try a closely
related symmetry-adapted non-orthogonal basis set given by
\begin{equation}
B^{S}=\left\{ P^{S}x^{k}y^{m}z^{n}\exp \left[ -a\left(
x^{2}+y^{2}+z^{2}\right) \right] ,\;k,m,n=0,1,\ldots \right\} ,
\label{eq:B^S_nonorth}
\end{equation}
where $a$ is a variational parameter that we set equal to unity for
simplicity. In order to obtain a suitable basis set $B^{S}$ it is necessary
to remove all the linearly dependent functions produced by the application
of the projection operator. Since this approach is quite straightforward for
programming we chose it for present calculations of basis set dimensions $%
N_{A_{1g}}=41$, $N_{E_{g}}=54$, $N_{T_{1g}}=66$, $N_{T_{2g}}=102$, $%
N_{A_{2u}}=16$, $N_{T_{1u}}=23$, $N_{T_{2u}}=39$. The first energy levels
are shown in Figure~\ref{fig:RROH} for $0\leq \lambda \leq 1$. Present
numerical results are accurate enough for the purpose of illustrating the
splitting of the degenerate energy levels of the harmonic oscillator as $%
\lambda $ increases. They are consistent with both the splitting
predicted by PGS (\ref{eq:splitting}) and the analytical
perturbation results (\ref {eq:PT_nu=0}-\ref{eq:PT_nu=3}). In
fact, we obtained those perturbation results from the secular
determinants obtained with the symmetry-adapted basis sets
(\ref{eq:B^S_nonorth}) with $a=1/2$ that is the exact value of
this parameter for $\lambda =0$. This strategy was already
described in Section~\ref {sec:perturbation}.

\section{Conclusions}

\label{sec:conclusions}

Throughout this paper we have tried to illustrate the application
of group theory to a quartic anharmonic oscillator with the
symmetry $O_{h}$ of the cube. We chose this particular example
because of its great symmetry and also because it was treated
before by means of a simpler symmetry-based approach consisting
only of parity and coordinate-permutation operators\cite{T89}. The
present application of group theory resorts to the 48 symmetry
operations shown in the table of characters in Table~\ref{tab:Oh}
and enables a systematic classification of the states of the
oscillator in terms of the corresponding irreps.

Group theory enables us to predict the splitting of the energy levels of the
harmonic oscillator as the perturbation parameter $\lambda $ increases. This
prediction is verified by the actual calculation of the perturbation
corrections. An advantage of the application of group theory is that we
treat each symmetry subspace independently of the others. In this way both
the calculation effort as well as the undesirably effect of the degeneracy
are considerably reduced. Note, for example that the degenerate energy
levels with $\nu =1$ and $\nu =2$ can be treated as nondegenerate when one
considers every symmetry species separately.

Group theory also enables us to reduce the dimension of the secular
equations in the application of the Rayleigh-Ritz variational method. In
addition to it, one obtains the eigenfunctions $\psi _{n}^{S}$ that are
bases for the irreps. This fact facilitates, for example, the calculation of
matrix elements of the form $O_{nm}^{SS^{\prime }}=\left\langle \psi
_{n}^{S}\right| O\left| \psi _{m}^{S^{\prime }}\right\rangle $ for any
observable $O$. Given the symmetry of $O$ one knows beforehand whether the
matrix element $O_{nm}^{SS^{\prime }}$ is zero\cite{H62,T64,C90}. This
analysis leads, for example, to the selection rules for molecular spectra%
\cite{H62,T64,C90}.

\begin{table}[tbp]
\caption{Character table for group $O_{h}$}
\label{tab:Oh}\scalebox{0.7}{\
\begin{tabular}{l|cccccccccc|l|l}
$O_h$ & $E$ & $8C_3$ & $6C_2$ & $6C_4$ & $3C_2(=C_4^2)$ & $i$ & $6S_4$ & $%
8S_6$ & $3\sigma_h$ & $6\sigma_d$ &  &  \\ \hline
$A_{1g}$ & 1 & 1 & 1 & 1 & 1 & 1 & 1 & 1 & 1 & 1 &  & $x^2+y^2+z^2$ \\
$A_{2g}$ & 1 & 1 & -1 & -1 & 1 & 1 & -1 & 1 & 1 & -1 &  &  \\
$E_g$ & 2 & -1 & 0 & 0 & 2 & 2 & 0 & -1 & 2 & 0 &  &
$(2z^2-x^2-y^2,
x^2-y^2) $ \\
$T_{1g}$ & 3 & 0 & -1 & 1 & -1 & 3 & 1 & 0 & -1 & -1 & $(R_x, R_y,
R_z)$ &
\\
$T_{2g}$ & 3 & 0 & 1 & -1 & -1 & 3 & -1 & 0 & -1 & 1 & $(xz, yz, xy)$ &  \\
$A_{1u}$ & 1 & 1 & 1 & 1 & 1 & -1 & -1 & -1 & -1 & -1 &  &  \\
$A_{2u}$ & 1 & 1 & -1 & -1 & 1 & -1 & 1 & -1 & -1 & 1 &  &  \\
$E_u$ & 2 & -1 & 0 & 0 & 2 & -2 & 0 & 1 & -2 & 0 &  &  \\
$T_{1u}$ & 3 & 0 & -1 & 1 & -1 & -3 & -1 & 0 & 1 & 1 & $(x, y, z)$ &  \\
$T_{2u}$ & 3 & 0 & 1 & -1 & -1 & -3 & 1 & 0 & 1 & -1 &  &
\end{tabular}
}
\end{table}

\appendix

\section{Simplified outline of Group Theory}
\label{sec:appendix}

In what follows we develop a simplified and abbreviated version of some of
the elements of group theory that are necessary for the present paper. A
rigorous account of group theory is available in any of the books on the
subject\cite{H62,T64,C90}.

Here we are interested in the analysis of the symmetry properties of a
physical system that are related to all the unitary operators $%
X_{i}^{\dagger }=X_{i}^{-1}$ that leave the Hamiltonian operator $H$
invariant
\begin{equation}
X_{i}HX_{i}^{\dagger }=H,  \label{eq:XiHXi+}
\end{equation}
and assume that this set is finite
\begin{equation}
G=\left\{ X_{1},X_{2},\ldots ,X_{h}\right\} .  \label{eq:G}
\end{equation}
The product (composition) of these unitary operators is
associative: $(X_{i}X_{j})X_{k}=X_{i}(X_{j}X_{k})$. The identity
operator $X_{1}$, which satisfies
\begin{equation}
X_{1}X_{i}=X_{i}X_{1}=X_{i},  \label{eq:X1}
\end{equation}
leaves the Hamiltonian invariant and, consequently, belongs to $G$. It
follows from (\ref{eq:XiHXi+}) that if $X_{i}$ belongs to $G$ then $%
X_{i}^{-1}$ belongs to $G$ too. The product of two operators also belongs to
$G$ as follows from
\begin{equation}
X_{i}X_{j}H(X_{i}X_{j})^{\dagger }=X_{i}X_{j}HX_{j}^{\dagger }X_{i}^{\dagger
}=X_{i}HX_{i}^{\dagger }=H.  \label{eq:XiXjH(XiXj)+}
\end{equation}
Because of all these mathematical properties the set $G$ is a
finite group.

Two such operators ( or group elements) $X_{j}$ and $X_{k}$ are said to be
conjugate if
\begin{equation}
X_{i}X_{j}X_{i}^{\dagger }=X_{k},  \label{eq:class_Xj,Xk}
\end{equation}
for some $X_{i}\in G$. If $X_{l}$ and $X_{k}$ are conjugate to $X_{j}$ then
they are conjugate to each other. All the mutually conjugated elements of a
group are collected into a class.

We can construct a matrix representation $\mathbf{X}_{i}$ of every operator $%
X_{i}$ in terms of a basis
\begin{equation}
B=\left\{ f_{1},f_{2},\ldots \right\} ,  \label{eq:group_basis}
\end{equation}
in the usual way
\begin{equation}
X_{i}f_{j}=\sum_{k}\left( \mathbf{X}_{i}\right) _{kj}f_{k},\;i=1,2,\ldots ,h.
\label{eq:Xifj}
\end{equation}
Of particular interest are the irreducible representations (irreps) obtained
in terms of suitable basis sets
\begin{equation}
B^{\alpha }=\left\{ f_{1}^{\alpha },f_{2}^{\alpha },\ldots ,f_{l_{\alpha
}}^{\alpha }\right\} ,\;\alpha =1,2,\ldots ,m,  \label{eq:B_alpha}
\end{equation}
such that
\begin{equation}
X_{i}f_{j}^{\alpha }=\sum_{k=1}^{l_{\alpha }}\left( \mathbf{X}_{i}^{\alpha
}\right) _{kj}f_{k}^{\alpha },\;j=1,2,\ldots ,l_{\alpha },\;i=1,2,\ldots
,h,\;\alpha =1,2,\ldots ,m.  \label{eq:Xi_fj^alpha}
\end{equation}
The characters of the irreps are the traces of the corresponding matrix
representations
\begin{equation}
\chi _{i}^{\alpha }=tr\left( \mathbf{X}_{i}^{\alpha }\right) .
\label{eq:chi_i^alpha}
\end{equation}
It can be proved that the number $m$ of irreps equals the number of classes
of group elements.

In order to obtain a basis function $f^{\alpha }$ for a given irrep we apply
the corresponding projection operator
\begin{equation}
P^{\alpha }=\frac{l_{\alpha }}{h}\sum_{i=1}^{h}\left( \chi _{i}^{\alpha
}\right) ^{*}X_{i},  \label{eq:P^alpha}
\end{equation}
to an arbitrary function $f$: $f^{\alpha }=P^{\alpha }f$. A projection
operator $P$ satisfies the following properties: $P^{\dagger }=P$ and $%
P^{2}=P$ ; more precisely, any operator that satisfies these two properties
is a projection operator. It follows from these properties that $%
\left\langle P^{\alpha }f\right| \left. P^{\alpha }f\right\rangle
=\left\langle f\right| \left. P^{\alpha }f\right\rangle \leq \left\langle
f\right| \left. f\right\rangle $.

In order to apply the equations above we should know how to
express the effect of a group operator $X_{i}$ on a function
$f(\mathbf{x})$ of the cartesian coordinates $\mathbf{x}=(x,y,z)$.
Rotations, reflections, etc can be expressed in matrix form as
$\mathbf{x}^{\prime }=\mathbf{Mx}$ so that there is a one to one
correspondence between the operators $X_{i}$ and the corresponding
transformation matrices $\mathbf{M}_{i}$. If we write
\begin{equation}
X_{i}f(\mathbf{x})=f\left( \mathbf{M}_{i}^{-1}\mathbf{x}\right) ,
\label{eq:Xif(x)}
\end{equation}
then we have the mappings
\begin{eqnarray}
\left( X_{i},X_{j}\right) &\rightarrow &\left( \mathbf{M}_{i},\mathbf{M}%
_{j}\right)  \nonumber \\
X_{i}X_{j} &\rightarrow &\mathbf{M}_{i}\mathbf{M}_{j}.  \label{eq:XiXj->MiMj}
\end{eqnarray}

It follows from (\ref{eq:XiHXi+}) that $[H,X_{i}]=0$ and, according to (\ref
{eq:P^alpha}), we conclude that $[H,P^{\alpha }]=0$.

In some cases it may be necessary to carry out an equivalent transformation
of the Hamiltonian of the system
\begin{equation}
UHU^{\dagger }=\tilde{H},\;U^{\dagger }=U^{-1}.  \label{eq:UHU+}
\end{equation}
The point group $\tilde{G}$ for $\tilde{H}$
\begin{equation}
\tilde{G}=\left\{ \tilde{X}_{1},\tilde{X}_{2},\ldots ,\tilde{X}_{h}\right\}
,\;\tilde{X}_{i}=UX_{i}U^{\dagger },  \label{eq:UGU+}
\end{equation}
is isomorphic with $G$ as follows from
\begin{equation}
\tilde{X}_{i}\tilde{X}_{j}=UX_{i}U^{\dagger }UX_{j}U^{\dagger
}=UX_{i}X_{j}U^{\dagger }=\widetilde{X_{i}X_{j}}.  \label{eq:UXiXjU+}
\end{equation}

Table~\ref{tab:Oh} shows the character table of the point group
$O_{h}$. The first row shows the symmetry operations grouped into
classes. $E$ is the identity operation, $C_{n}$ denotes a rotation
by an angle $2\pi /n$, $i$ is the inversion operation, $S_{n}$
denotes a rotation by an angle $2\pi /n$ followed by a reflection
with respect to a plane perpendicular to the rotation axis and
$\sigma $ indicates a reflection plane. The first column displays
the irreps; those labelled by $A$, $E$, and $T$ are one-, two- and
three-dimensional (that is to say, $l_{\alpha }=1,2,3$),
respectively. The numbers are the characters $\chi _{i}^{\alpha }$
and the last two columns show some basis functions for the irreps.
This table summarizes some of the relevant ingredients for the
construction of the projection operators $P^{\alpha }$ shown in
equation (\ref{eq:P^alpha}). The matrix representation of the
elements of this group, which is necessary for the construction of
the symmetry operations (\ref{eq:Xif(x)}) and projection operators
(\ref {eq:P^alpha}), is available elsewhere\cite{F14}.

\end{document}